\documentstyle[12pt]{article}
\def\Cerenkov{\v Cerenkov }

\begin{document}
\baselineskip=0.7cm

\begin{flushright} Preprint USM-TH-66 \end{flushright}
\vspace*{.2in}

\begin{center} \Large\bf Antihydrogen\end{center}
\vspace*{.10in}

\begin{center}
Iv\'an Schmidt\footnote[1]{Work supported in part by FONDECYT 
(Chile), 
contract 1960536}   \end{center}
\begin{center}
{\it Department of Physics \\ Universidad Federico Santa 
Mar\'\i a \\
Casilla 110-V \\ Valpara\'\i so, Chile}  
\end{center}
\vspace*{.10in}
\begin{abstract}
\baselineskip=0.4cm

{\footnotesize 
CERN announced in January 1996 the detection of the first eleven 
atoms of antimatter ever produced. The experiment was based on a method 
proposed earlier by S. Brodsky, C. Munger and I. Schmidt, and which 
furthermore predicted exactly the number of atoms that were detected
for the particular conditions of the experiment. The study of 
antihydrogen affords science the opportunity to continue research on the 
symmetry between matter and antimatter. In this talk the importance of 
antihydrogen as a basic physical system is discussed. Different 
production methods that have been tried in the past are briefly 
presented, and the method that was used in the CERN experiment is 
analyzed in detail. It consists in producing antihydrogen by 
circulating a beam of an antiproton ring through an internal gas 
target. In the Coulomb field of a nucleus, an electron-positron pair 
is created, and antihydrogen will form when the positron is created 
in a bound rather that a continuum state about the antiproton. The 
theoretical calculation of the production cross section is presented 
in detail. A discussion of the detection systems used both in the 
CERN experiment and in another similar experiment that is right now 
underway at Fermilab  are also given. Finally I present and discuss   
possible future experiments using antihydrogen, including the 
measurement of the antihydrogen Lamb shift. }

\end{abstract}
\vspace{.5in}

To appear in the proceedings of the First Latin American Symposium on 
High Enery Physics (SILAFAE), M\'erida, M\'exico, October 
30-November~6, 1996.

\newpage
\baselineskip=0.7cm 
\parskip=9pt

Antihydrogen, the simplest atomic bound state of antimatter, 
$\bar H \equiv (e^+\bar p)$, had never been observed experimentally 
until the CERN announcement of January 1996: a team of physicists had
succeeded for the first time in synthesizing atoms of antimatter from 
their constituents antiparticles [1]. The particular method 
that was used had been suggested first in 1990 [2], and the 
detailed calculation presented later [3,4]. The news of     
this discovery appeared in all major newspapers and scientific        
journals in the world. 
 
\section{Why antihydrogen ?}

The hydrogen atom has been one of the most important physical systems 
for a wide variety of fundamental measurements related to the behavior 
of ordinary matter. The production of antihydrogen opens the door for 
a systematic exploration of the properties of antimatter, and give us 
unique possibilities of testing fundamental physical                  
principles[5].

\paragraph*{\bf CPT Invariance:~}
The CPT theorem is a basic result of quantum field theory, and it is  
the essential ingredient for knowing the properties of antimatter.
CPT invariance implies that a particle and its antiparticle have:     
equal and opposite additive quantum numbers, such as electric charge, 
and equal masses, total lifetimes and gyromagnetic ratios.
Nevertheless, CPT violation could occur when non-finite-dimensional 
representations of the Lorentz group are permitted, in curved 
spacetimes, or through non-localities.

In the case of antihydrogen, CPT invariance requires that the its     
spectrum has to be exactly equal to that of the hydrogen atom. Since  
the hyperfine splitting in hydrogen is known to a precision of         
roughly $6~ 10^{-12}$, comparison with antihydrogen will provide      
a superlative direct CPT test for baryons. 

\paragraph*{\bf Equivalence principle for antiparticles:~}
The question is whether antimatter might behave differently than 
ordinary matter in a gravitational field. Since
transition frequencies are subject to gravitational red shifts 
when the system is moved through a gravitational potential difference, 
this could be the basis of experimental tests of the equivalence 
principle for antimatter. 

\section{Production}

An important obstacle for the production of antihydrogen is the high 
energy of the antiproton beam. By using special devices such as       
Penning traps, researchers have been able to reduce substantially 
their energy. The next problem is to combine the antiprotons with 
positrons. Several sources are currently under development. For 
example [6]:

\paragraph*{\bf Radiative recombination},
$\overline {p}+e^{+}\rightarrow \big(\overline {p}e^{+}\big)+\gamma$. 
~The problem is that the collision time ($\sim 10^{-15}s$) is much     
less than the typical radiative lifetime ($\sim 10^{-9}s$). Thus it   
takes a long time to radiate a photon compared to the collision time.

\paragraph*{\bf Laser assisted Recombination},  
$\overline {p}+e^{+}+n\gamma \rightarrow \big(\overline
{p}e^{+}\big)+(n+1)\gamma $.~  
Unfortunately the verified gain is high only for states with $n > 8$.

\paragraph*{\bf Plasma Recombination}, 
$\overline {p}+e^{+}+e^{+}\rightarrow \big(\overline {p}e^{+}\big)+e^{+}$.~
Three way collision, in which the extra particle would carry off the 
excess energy.
      
\paragraph*{\bf Capture from Positronium},
$\overline {p}+\big(e^{+}e^{-}\big)\rightarrow \big(\overline
{p}e^{+}\big)+e^{-}$.~  
 
\paragraph*{\bf From antiprotonic helium states},
$\big(\overline {p}\overline {He}\big)+\big(e^{+}e^{-}\big)\rightarrow
\big(\overline {p}e^{+}\big)+\big(e^{-}He\big)$.~
The problem here is the annihilation of the antihydrogen within the 
liquid helium.
  
No antihydrogen has yet been made by any of these processes, yet curiously 
a working source had been producing antihydrogen for some time, 
but with the antihydrogens going out undetected. This is the process 
that was used in the CERN experiment, colliding
relativistic antiprotons with ordinary atoms.
An antiproton passing
through the Coulomb field of a nucleus of charge $Z$ will create
electron-positron pairs; occasionally a positron
will appear in a bound instead of a continuum state about the moving
antiproton to form antihydrogen.
The cross section for this capture process
we calculate to be
$\sigma(\bar p Z \to \bar H e^- Z) \sim 4~Z^2 pb$
for antiproton momenta above $\sim 6$ GeV/c.
 
\section{\bf Cross Section Calculation} 

The production process for antihydrogen studied here
is the exclusive two- to three-particle reaction
$\bar p p \to \bar H e^- p'$.
The equivalent photon approximation [7],
applied in the antiproton rest frame, relates
the cross section for pair creation with capture to
$1s$ state,
$\bar p  Z
\rightarrow \bar H(1s)  e^-  Z$,
to the cross section for photon-induced capture using virtual photons,
$\gamma^*  \bar p \to e^-  \bar H(1s)$.
Consistent with this approximation we assume the $\bar p$
and the $\bar H$ in these processes to remain at rest in a
common frame of reference, and neglect the electron and photon
energies compared with the antiproton mass.
We find
\begin{eqnarray} 
&&\sigma_{\bar p Z\rightarrow\bar H(1s)\, e^- Z}=
  {Z^2\alpha\over\pi}\,\,\times \cr
&&\quad
\int^1_{2m/E}{dx\over x}
\int^{q^2_{\bot max}}_0 dq^2_\bot {q^2_\bot\over
(q^2_\bot+x^2 M^2)^2}\
\left(1+(1-x)^2\over2\right)\,
\sigma_{\gamma^*\bar p\rightarrow\bar H(1s) \,  e^-}(\omega,q^2).
\end{eqnarray} 
Here $x = {\omega/ E} =
{ \omega / (M \gamma)}$ is the photon energy
fraction evaluated in the antiproton rest frame,
$q_\bot$ is the photon's transverse momentum, and
$M$ is the antiproton mass.
At large photon virtuality $Q^2= - q^2$
the  photoabsorption cross section falls off as
$(Q^2+ 4 m_e^2)^{-1}$, so the transverse momentum
$q_\bot$ in the integrand is typically of order $2 m_e$.
The upper limit of integration $q_{\bot max}$ has
the same order of magnitude as the photon energy $\omega$
because
the photon tends to have $q^2\approx 0.$
 
The contributions where $q^2$ is small dominate the integral
and so we can set $q^2 \approx 0$
and substitute the cross section
for real instead of virtual photons.
Performing the integral over $q_{\bot}^2$ we find
that the total cross section factors,
\begin{equation}
\sigma_{\bar p  Z\rightarrow\bar H(1s) \, e^- Z}(\gamma)
 = {Z^2} F(\gamma)  \, {\alpha \over \pi} \bar\sigma,   
\end{equation}
where
\begin{equation}
F(\gamma)=\ln(\gamma^2 + 1)-{\gamma^2 \over \gamma^{2} +1},
\end{equation}
contains the dependence on the relative velocity of the
antiproton and the nucleus, where $\gamma$ is the Lorentz factor
$\gamma = (1-\beta^2)^{-1/2}$, and where
\begin{equation}
\bar\sigma=\int_{2m_e}^E {d\omega \over \omega}
\sigma_{\gamma\bar p\rightarrow\bar H(1s) \,  e^-}(\omega)
\end{equation}
is a constant.
 
Crossing symmetry relates the matrix elements for
photon-induced capture and for photoionization:
the sum over initial and final
spins of the squares of the matrix elements for
the reactions $\gamma \bar p \rightarrow e^- \bar H(1s)$ and
$\gamma \bar H(1s)\rightarrow e^-  \bar p$ become equal if
they are written in terms of the Mandelstam variables
$s$, $t$, and $u$,
and if the variables $s$ and $u$ in one element are exchanged.
The squares of the
matrix elements when the final spins are
summed but the initial spins are averaged, $|M|^2$,
are likewise related except that a factor of two is introduced:
\begin{equation}
|M|^2_{capture}(s,t,u) = 2 |M|^2_{photo}(u,t,s).
\end{equation}
The matrix element for photoionization is well known [7]
so we easily obtain the matrix element for capture.
In the $\bar p$ rest frame the Mandelstam variables for
photoionization are equal to
\begin{eqnarray}
s=& M^2+2M\omega, \cr
t=& m^2-2\omega\epsilon + 2 \vec k \cdot \vec p,  \cr
u=& M^2-2\epsilon M+m^2,
\end{eqnarray}
where
$\vec{p}$ and $\vec{k}$ are respectively
the momentum vectors of the outgoing electron
and incoming photon, and $\epsilon = \sqrt{|\vec p\,|^2 + m^2}$
is the electron energy.  We define also $p = |\vec p\,|$
and $k = |\vec k| = \omega$.
 
After rewriting the matrix element for photoionization
in terms of the
Mandelstam variables, and exchanging $s$ and $u$,
we find in
the $\bar p$ rest frame that 
\def\fixit{\vphantom{(\vec k - \vec p)^2}} 
\begin{eqnarray}
\ |M|^2_{capture} =
{64 \pi^2 \alpha^5 m^4 \over \epsilon (\vec{k}-\vec{p}\,)^4}
\Biggr[&
{a^2p^2m\over \fixit \epsilon + m}
+ a(\vec k \cdot \vec p - p^2)
  \biggl( {1\over \fixit k^2 - p^2}
         + {1\over (\vec k - \vec p\,)^2} \biggl)\hfill \cr
&+ {\epsilon + m \over \fixit 4 m} (\vec k - \vec p\,)^2
  \biggl( {1\over \fixit k^2 - p^2}
           + {1\over (\vec k - \vec p\,)^2} \biggr)^{\!2} \cr
&- {\epsilon + m \over \fixit 2m}
  {k^2p^2 - (\vec k \cdot \vec p\,)^2 \over
   (k^2-p^2) (\vec k - \vec p\,)^2 k^2     }
+ a
  {k^2p^2 - (\vec k \cdot \vec p\,)^2 \over
   (\vec k - \vec p\,)^2 k^2     }
\Biggr]
\end{eqnarray}
where the quantity $a$ is given by
\begin{equation}
a={1\over (\vec k - \vec p\,)^2}
+ {\epsilon \over \fixit m (k^2-p^2)}.
\end{equation}
The total cross section for photon-induced capture is
\begin{equation}
\sigma_{\gamma  \bar p \to e^-  \bar H(1s)}(\omega) =
{\alpha \epsilon p \over 2 \pi M^2} \int
|M|^2_{capture} d\cos\theta.
\end{equation}
 
Integrating the previous equations numerically
yields $(\alpha/\pi)\bar\sigma = 1.42~{\rm pb}.$
We then get the cross section for
$\bar p Z \to \bar H(1s) e^- Z$ as a function of the antiproton
momentum; it is approximately $4Z^2$ pb for
momenta above {$\sim 6$ GeV/c}.
Capture into states of higher principal quantum number
will increase the total cross section for capture by
$\sim 10\%-20\%$.

In the case of the CERN experiment, the momentum of the antiproton is 
$1.94~GeV/c$, which gives a cross section of approximately $2\ pb\ \times
\ Z^{2}\approx 6\ \times \ 10^{-33}cm^{2}$. With an integrated        
luminosity of $5\ \times \ 10^{33}cm^{-2}\pm 50\%$, and an acceptance 
of $\epsilon =0.3$ for the detection system, the expected number of   
events is about $10$. The actual number was $11$ ! 
Right now  a similar experiment is underway at Fermilab. As of this 
writing they have detected $18$ antihydrogen atoms [8]

Calculations of similar cross sections
have been reviewed by Eichler [9], and
give $\sigma(\bar p p \to \bar H p e^-)
= 2.7 \ln(\gamma)~{\rm pb}$,
which is very close to our asymptotic result
$\sigma = 2.8 \ln(\gamma)~{\rm pb}$.
Becker [10] computes the
cross section for two different momenta; the ratio of
the cross sections agrees
remarkably well with our prediction,
though his cross sections are lower than ours by a factor of $2.8$.

\section{Detection}
    
The momentum transferred to the antiproton
in the process $\bar p Z \rightarrow \bar H  e^-  Z$
is small, the order of
$m c \sim 5\cdot 10^{-4} GeV/c$
The momentum and position vectors
of an antihydrogen atom are therefore
the same as those of the antiproton
from which it forms;
a monoenergetic, small-divergence
bunch of antiprotons exits a cloud of gas
overlapped
by an equivalent bunch of antihydrogen atoms.

It can easily be shown that the antihydrogen atoms escape the gas 
intact, and that it easily survives
its escape through the dipole fields of the Accumulator
ring [4]. Fast antihydrogen
separates into a pair of free particles
with a probability greater than 0.99 in a mere membrane of 
polyethylene $400\ \mu{\rm gm}\,{\rm cm}^2$ thick.
An antihydrogen atom therefore
generates in coincidence, from some point in
a known, few-square-centimeter area
of a membrane possibly tens of meters from the gas target,
a positron and an antiproton with a common and tightly constrained
velocity equal to the known
velocity of the antiprotons circulating in the storage ring.
So spectacular is this signature that the chief difficulty in
designing an apparatus to detect antihydrogen
is to choose which of many sufficient schemes is the simplest.

The CERN experimental detection system is roughly as follows. As a 
neutral object the antihydrogen, produced ten meters upstream in the 
center of the straight section of LEAR, will exit the accelerator 
ring tangentially and will be stripped in the first silicon          
counter. Then the resulting antiproton and positron hit simultaneously 
this first silicon counter which, together with a second one, measures the 
$dE/dx$ of the antiproton plus the kinetic energy of the positron 
being stopped in these two detectors. The third silicon counter should 
give a signal proportional to the $dE/dx$ of the $\overline {p}$ 
only. The two $511\ KeV$ photons from the $e^{+}-e^{-}$ annihilation  
are detected back to back in a NaI counter. The silicon counter       
telescope is located at the center of the NaI crystal arrangement. 
The $\overline {p}$'s resulting from the stripped $\overline {H}$     
continue with a velocity of $\beta =0.900$. They penetrate through a  
set of three start scintillators, a scintillator hodoscope and a       
group of stop scintillators. The deflection of the registered charged 
particles was measured with a spectrometer, and it should be within 
the predicted values for a $\overline {p}$ from $\overline {H}$.
Several cuts were performed to the data in order to reach the         
conclusion that $11$ $\overline {H}$ had been observed.

In the case of the Fermilab experiment the detection system is        
slightly different.
An antihydrogen atom exits the Accumulator ring
and strikes a 400 $\mu{\rm gm} \, {\rm cm}^{-2}$ membrane, where it
separates into a positron and an antiproton.  Because the particles'
magnetic rigidities differ by a factor of
nearly 2000
the positron can be bent away
and focussed while hardly affecting the antiproton,
so the particles can be directed into separate detectors.
We describe how these detectors function for an antihydrogen
momentum of
3 GeV/c; the apparatus works equally well up to the maximum
Accumulator momentum of 8.8 GeV/c merely by scaling various
magnetic fields.
 
A positron spectrometer a few meters in length,
consisting in succession of a solenoid lens, a sector magnet,
and a second solenoid lens, separates the positrons and
focuses them onto a few square-centimeter
spot. A momentum resolution of a few percent
matches the Fermi smear in the momentum of the positrons.
The positrons stop in a scintillator $\sim 1$ centimeter
thick whose
total volume is only a few cubic centimeters.
Light from
the scintillator is guided into a single phototube;
the rise  and height of the output pulse
give respectively
a measure of the arrival time of the positron within 1 ns, and a measure
of its kinetic energy to 20\%.
A $4\pi$ NaI detector surrounds
the scintillator and
intercepts the two
511-keV photons from the positron's annihilation;
the detector also vetoes the passage of stray charged particles,
which deposit far more energy than 511 keV.
The collection efficiency of the spectrometer is 99\%.
Some 95\% of the positrons come to rest in the scintillator and deposit
their full energy; the 5\% that backscatter nonetheless deposit
enough energy to make a signal, and still come to rest and annihilate
in the volume surrounded by NaI.
 
Tagging the antiproton is straightforward.  Over a flight distance
of 40 meters it passes through a pair of scintillator paddles
that measure its velocity by time of flight; a sequence of
multiwire proportional chambers (MWPC's) and bend magnets that
measure its momentum to 0.2\%, and a terminal \Cerenkov
threshold detector that provides a redundant velocity separation
of antiprotons and other negative particles.  The tiny beamspot,
angular divergence, and unique and known
velocity of the antiproton favor a
differential-velocity
\Cerenkov detector , but for momenta as low as
3 GeV/c it is difficult to generate enough light for such a device
without using a \Cerenkov medium so thick that antiprotons do not
survive their passage.

A candidate antiproton is defined by
hits in the
MWPC's consistent
with a 3~GeV/c
antiproton that originates in the right
few square-centimeter area of the membrane, by
the absence of a hit in the \Cerenkov threshold detector,
and by
hits in the time-of-flight
scintillators consistent with the passage of the
particle of the right velocity.  A candidate positron is defined
by a hit in the positron scintillator with the right deposit
of kinetic energy, coincident within a few nanoseconds with
a hit in the surrounding NaI
consistent with the absorption or Compton scatter of two
511 keV photons.
An antihydrogen candidate is defined as a sub-nanosecond coincidence
between antiproton and positron candidates.
 
\paragraph*{\bf Backgrounds:}~
Despite the $\sim 10^{10}$
times higher cross section for $\bar p p$ annihilation than for
antihydrogen production, the backgrounds from particles
originating in the gas target
will be zero.  Here we consider only those processes which have such
peculiar kinematics and large branching ratio as to mimic at least
part of an antihydrogen signal, most importantly the antiproton,
without requiring a failure of some part of the detector. We will 
consider just the Fermilab experiment, but a very similar analysis    
applies to the CERN experiment. 
 
The only particle which can satisfy the momentum
and velocity measurements made by the particle tracking and by the
time-of-flight and
\Cerenkov detectors is an antiproton.  Antiprotons lost from
the ring are unimportant; few thread properly through the target and
the wire chambers, and the passage of an antiproton through even
an extra 2 cm of aluminium will slow the antiproton below the
0.2\% momentum resolution of the particle tracking.
The only plausible source of antiprotons is from antineutrons,
made in the target by
the charge-exchange reaction $\bar p p \rightarrow \bar n n$,
that convert by a second charge-exchange reaction into
an antiproton.
The simplest possibility is for an antineutron generated in
the gas target to pass undeflected through the Accumulator's
dipole magnet
and convert in the first MWPC;
the new antiproton fools the rest of the detectors in our
antiproton beam line.
Of all processes that make antiprotons this one has
the most dangerous combination of unfavorable kinematics and
big relevant cross sections.
The number of $\bar n$'s made
in the gas target is large; both the neutral $\bar n$'s and the
converted $\bar p$'s are thrown forward, the more easily
to pass through
the apparatus and the antiproton beamline;
and the forward $\bar p$'s from this two-step process can have
precisely the same velocity as the antiprotons circulating in the
ring.
 
Nonetheless the beamline will count fewer antiprotons
from this process than it will count antiprotons from antihydrogen.
The count rate is small principally
because of the small
solid angle subtended
by the known
small spots (of order $3 \ {\rm cm}^2$)
that legitimate antiprotons make on both
the first and last MWPC's, compared to the typical solid angle
over which particles from charge exchange are distributed.
At a model momentum of $p = 3$ GeV/c
the cross section for $\bar p p \rightarrow
n \bar n$ is 2.0 mbarn; an integrated luminosity of
\def\pb{${\rm pb}^{-1}$}
200 \pb~ will produce $4.0\cdot 10^{11}$ antineutrons.  In charge
exchange the typical momentum transfer is the order of the pion
mass, {$\delta p \sim 135$ Mev/c},
and so the outgoing antineutrons will be distributed
over a solid angle that is the order of
$\Delta \Omega \sim \pi (\delta p/p)^2 \sim
6.4\cdot 10^{-3}$ sterradian.  The thickest material in front
of all the MWPC's is 0.3 cm of scintillator in the first
time-of-flight detector. Reconstruction of the antiproton track
limits the active area on this scintillator to roughly
3 ${\rm cm}^2$; the solid angle of this spot as seen from the
gas target, 20 meters away, is only
$7.5\cdot 10^{-7}$
sterradian, and the probability any antineutron from charge
exchange hits the spot is only $1.2\cdot 10^{-4}$.
The probability
of a hadronic interaction in the scintillator is only $3.8\cdot 10^{-3}$;
if an antineutron
interacts, it will generate an antiproton with a probability
equal to the ratio of the cross section for charge exchange to
the total cross section, which for a proton target is equal
to $2.0 {\ \rm mbarn} / 79.9{\ \rm mbarn} = 2.5\cdot 10^{-2}$.
Any antiproton made must now pass though the antiproton beam line and
strike a 3 ${\rm cm}^2$ spot on the
last MWPC, at least 20 meters away; because of the
momentum transfer in charge exchange the probability it will do so
is again only $1.2\cdot 10^{-4}$.  Collecting all the factors, the
number of antiprotons from this process indistinguishable from
the 760 antiprotons we expect from the separation of antihydrogen is
at most
$$4.0\cdot 10^{11} \times 1.2\cdot 10^{-4} \times
3.8\cdot 10^{-3} \times 2.5\cdot 10^{-2} \times 1.2\cdot 10^{-4}
= 0.5$$
This order-of-magnitude argument is quite crude but sufficient to
show that the rate of such false antiprotons, occurring in coincidence
with a positron candidate, is certainly negligible.

The kinematics of an antineutron' ordinary
$\beta$-decay, $\bar n \rightarrow \bar p e^+ \nu$,
are also unfavorable because the decay
produces not only a 3 GeV/c antiproton but also
a positron whose range of laboratory kinetic energy
overlaps 1.200 MeV; fortunately its slow decay rate
$\sim 10^{-3} \, {\rm s}^{-1}$ prevents a significant fraction from
decaying within our apparatus.

\section{Other experiments with relativistic antihydrogen}
 
To test CPT invariance it is best
to study hydrogen and antihydrogen in the same
apparatus.
The polarity of the magnets in the Fermilab
Accumulator can be reversed and protons circulated; as
the protons pass through the target gas they pick up atomic electrons
and make a neutral hydrogen beam that has the same optics as
the antiproton beam.  For protons above 3 GeV/c
the dominant process
is one in which an essentially free electron in the target falls
into the $1s$ state and a photon carries off the binding energy.
The cross section per target electron
is $\geq 1.7$ nanobarn, so for equal circulating currents through
a hydrogen
gas target the hydrogen beam will have $\sim 430$
times
the intensity of the antihydrogen beam.
 
Two experiments seem practical with meager
samples respectively
of order $10^3$ and $3\cdot 10^4$ antihydrogen atoms.
The first is a measurement of
the rate of field ionization of
the $n=2$ states in an electric field provided by the Lorentz
transform of a laboratory magnetic field.
Roughly 10\% of a
$3 GeV/c$
beam of antihydrogen
in the $1s$ state can be excited into
states with $n=2$ by passing it
through a thin membrane.
If the membrane sits in a 20 kgauss transverse magnetic
field, equation shows that
the states with $n>2$ will ionize instantly, that the states with
$n=2$ will
ionize with $1/e$ decay lengths of order 10 cm, and that the $1s$ state
will not ionize at all.  The distance a state with $n=2$ flies
before ionizing is marked by
the a deflection
of the freed antiproton by the magnetic field
by an amount
between the zero deflection of the surviving $1s$ component of
the beam and the large deflection of the antihydrogen that
ionizes instantly or separates in the membrane.
Ten centimeters of flight before ionization changes the
deflection of the antiproton seen 3 meters away by
6.7 cm, many times the antiproton spot size
of $\leq 1 cm$; and  changes the antiprotons angle by 22 mrad, 
many times both its original
angular divergence of 0.2 mrad and the
resolution of 0.1 mrad that can be provided by a pair of
MWPC's with
{1 mm} resolution and 10 meters apart.
The distance a state flies is also marked by the freed positron,
whose orbit radius is only 2 mm in the transverse field, and which
can be directed along the field lines
into some sort of position-sensitive detector.
The positron and antiproton have of course their usual known
common velocity.
A flux of a few thousand $\bar H$'s may be sufficient to measure
the three distinct field ionization rates of the $n=2$ states
to $\sim 10\%$.  Because ionization is a tunneling
process its rate is surprisingly sensitive to details of the antihydrogen
wavefunction; a $10\%$ shift would require for example a change in
$\left<r\right>$ for the $n=2$ states
of only $0.24\%$.
 
Evidently
ionization in a magnetic field can be used to count efficiently
states with $n = 2$ without counting states of different
principal quantum number.
No other method is available; the Accumulator runs with long
antiproton bunches, and
no laser for example
has sufficient continuous power to photo-ionize
efficiently the relativistic antihydrogen beam.
By driving the 1000 Mhz $2s-2p$ transition and monitoring the
surviving $2s$ population as a function of frequency
the frequency of
the antihydrogen Lamb shift can be measured.
The $1/e$ decay length of the $2p$
states is 1.4 meters at 3 GeV/c,
so a few meters from an excitation membrane
only the
$2s$ population will survive.
The Doppler-shifted
transition can be driven
by chasing the beam with
6.1 GHz radiation aimed down
a waveguide roughly that is
10 meters long and roughly 10 ${\rm cm}^2$ in cross section;
this guide may also
serve as a beam pipe.
The $2s-2p$ resonance
has a quality factor of only 10
because of the width of the $2p$ state,
and so the Doppler broadening of
the transition, or a misalignment of the axes of the beam and the
guide,
will not put the transition out of resonance.
Because the $2s-2p$ transition is electric dipole in character
and has a large matrix element,
modest laboratory powers
of roughly 10 {{\rm Watt}/$cm^2$} suffice
to mix
the $2s$ state
completely with the $2p$ and make the $2s$ state
decay with a $1/e$ distance
of $2.8$ m.
To prevent
Stark mixing of the $2s$ and $2p$ states,
transverse
magnetic fields must be less than
0.1 gauss from the excitation membrane down the guide's length
until the sharp rise of the
transverse magnetic field that is
used to ionize and count the $n=2$ states.
If the rise occurs over less than the fully mixed $2s$ decay length
of 2.8 meters little of the $2s$ state will decay to the $1s$
instead of ionize.
A sample of a 300 antihydrogen atoms in the
$2s$ state would suffice to see a dip in the transmitted $2s$
population as a function of drive frequency, find its center
to within 10\% of its width, and so measure
the antihydrogen Lamb shift
to $\sim 1$\%.  A total of $3\cdot 10^4$ antihydrogen atoms is enough
to provide such a sample
if a membrane yields
as expected
$0.01$ $2s$ states per incident $1s$.
The experiment would be sensitive to a
differential shift of the $2s$ and
$2p$ states of hydrogen and antihydrogen equal to a
fraction $\sim 2\cdot 10^{-8}$ of the states' binding energy,
and would test the CPT symmetry of the $e^+\bar p$ interaction
at momentum scales characteristic of atomic binding, 10 eV/c.

\end{document}